\documentstyle[preprint,aps]{revtex}
\begin{document}
\draft
\preprint{}
\title{``Electromagnetic'' Contribution to the Nucleon Spin}
\author{D. Singleton}
\address{Department of Physics, Virginia Commonwealth University, 
Richmond, VA 23284-2000}
\date{\today}
\maketitle
\begin{abstract}
A possible resolution to the question of where the nucleon gets
some portion of its internal angular momentum, not carried by
the valence quarks, is given in terms of the electromagnetic field
angular momentum of the nucleon. This field angular momentum
is similiar in origin to that of the Thomson dipole (an electric charge
and a magnetic monopole), but arises from the interaction of the
quark electric charges and magnetic dipoles. Rough, order of 
magnitude calculations 
show that such a field angular momentum could provide 
some of the nucleon's spin. Under assumptions similiar to those
used to calculate hadron mass splittings, QCD should
also exhibit a similiar field angular momentum coming from the 
interaction of color charges with color magnetic dipoles.  
\end{abstract}
\newpage
\narrowtext

\section{Introduction}
In the simple quark model the spin 1/2 nucleon would get its spin 
from the three valence quarks of which it is constructed.
Starting with the European Muon Collaboration experiments \cite{emc}
it has been realized that the valence quarks are not the only
contributors to the total spin of the nucleon. The source of the
extra internal angular momentum of the nucleons is taken as coming
from virtual quark-antiquark pairs within the nucleon (mainly
the strange quarks), or from the gluons. The spin 1/2 of the
nucleon can be written out as
\begin{equation}
\label{spin}
{1 \over 2} = {1 \over 2} \Delta \Sigma + L_q + L_g + \Delta G
\end{equation}
where $\Delta \Sigma /2$ is the quark spin contribution, $L_q$ and
$L_g$ are the quark and gluon orbital contributions respectively, 
and $\Delta G$ is the gluon spin contribution. In this paper we
suggest that the contribution from photons should also be included,
and we show how a field angular momentum can result from the interaction
between the magnetic dipole moments and electric charges of the 
valence quarks. Under the assumption that this electrodynamic
angular momentum has at least a 
qualitative counterpart in QCD, we argue that a similiar
field angular momentum should arise from the interaction between 
the color magnetic dipole moments and color charges of the valence
quarks.

\section{Dipole-Charge Field Angular Momentum}

To find the field angular momentum contained in the configuration of
a magnetic dipole and an electric charge the angular momentum field
density should be integrated over all space.
\begin{equation}
\label{fangmom1}
{\bf L}_{em} = {1 \over 4 \pi } \int {\bf r} \times ({\bf E} \times
{\bf B}) d^3 x 
\end{equation}
${\bf E} = e {\bf r'} /{ r'} ^3$ is the Coulombic electric field
of the charged particle, $+e$, located at ${\bf R}$ (so ${\bf r'} = {\bf r}
- {\bf R}$), and ${\bf B} =  (3 {\bf r} ({\bf r \cdot M}) / r^5) -
({\bf M} / r^3)$ is the magnetic field of the dipole, ${\bf M}$, located 
at the origin. A quick method to find ${\bf L}_{em}$ directly is to
use the result that the field angular momentum of a magnetic charge-
electric charge system is ${\bf L} _{em} = e g {\bf {\hat r}}$, where
$+e$ and $+g$ are the electric and magnetic charges respectively, and
${\bf {\hat r}}$ is a unit vector which points from the electric charge
toward the magnetic charge \cite {jackson}. Placing a magnetic charge
of $+g$ at the origin, a magnetic charge of $-g$ at the point ${\bf
P}$, and an electric charge of $+e$ at ${\bf R}$ yields a field
angular momnetum of
\begin{equation}
\label{fangmom2}
{\bf L}_{em} = e \left( -{g {\bf R} \over \vert {\bf R} \vert}
+ {g ({\bf R - P}) \over \vert {\bf R - P} \vert} \right)
\end{equation}
Expanding $1 / \vert{\bf R - P} \vert$
and taking the limit ${\bf P} \rightarrow 0$, 
$g \rightarrow \infty$, and $g {\bf P} \rightarrow - {\bf M}$ gives
\begin{equation}
\label{amdipole}
{\bf L} _{em} = {e {\bf M} \over  R} - {e {\bf R} ({\bf M}
\cdot {\bf R}) \over  R^3}
\end{equation}
for the field angular momentum in the dipole-charge system. 
Directly integrating Eq. (\ref{fangmom1}) for this system
again yields the result in Eq. (\ref{amdipole}). Unlike the
field angular momentum of the magnetic charge-electric charge 
system, which is independent of the distance between the charges,
the dipole-charge field angular momentum has a $1 / R$
dependence. Thus as one shrinks the separation between the
dipole and the charge this field angular momentum will become
increasingly important. 

One subtle point with this derivation of the dipole-charge 
field angular momentum is that the magnetic dipole 
produced by two monopoles (a Coulombic dipole) is physically
different from the magnetic dipole produced by currents (an Amperian
dipole). For an Amperian dipole produced by a small current loop in
the xy-plane with the current flowing counterclockwise, the magnetic
field near the origin will point along the positive z-axis. For a
Coulombic dipole to produce the same kind of magnetic field away
from the origin, as the previous Amperian dipole, one should place
a $+g$ magnetic charge along the positive z-axis and a $-g$ magnetic 
charge along the negative z-axis. Now, however the magnetic field
of the Coulombic dipole will point along the negative z-axis near
the origin. Mathematically this is taken into
account by adding a delta function term to the standard magnetic dipole
field. One adds $+ 8 \pi {\bf M} \delta ({\bf r}) / 3$ for the magnetic
dipole produced by currents, and $- 4 \pi {\bf M} \delta ({\bf r}) / 3$
for the magnetic dipole produced by monopoles. The limiting procedure
for obtaining Eq. (\ref{amdipole}) only produces the $(3 {\bf r
(r \cdot M)} / r^5) - ({\bf M} / r^3)$ part of the magnetic dipole
field, and not the delta function part. Thus for each of the
two cases one should add to Eq. (\ref{amdipole}) the contribution
of this delta function part of the magnetic dipole field. In both
cases however the angular momentum density coming from the delta
function part of the magnetic field will be of the form
\begin{equation}
{\bf r} \times ({\bf E} \times {\bf M} \delta ({\bf r}))
\end{equation}
When this is integrated over all space it will vanish because of
the delta function and the factor of ${\bf r}$ in the integrand.
Thus Eq. (\ref{amdipole}) is the expression for the field angular 
momentum for either a magnetic dipole produced by currents or by 
monopoles.

\section{Eletromagnetic Field Angular Momentum of the Nucleon}

We now give a rough estimate of the angular momentum contributed
to the nucleon by the electromagentic field, to show that it may be
possible that some of the extra internal angular momentum of the 
nucleon may have an electromagnetic origin. We will consider the
proton although similiar considerations apply to the neutron. The
state vector for the proton can be written as \cite{don}
\begin{equation}
\label{proton}
\vert p _{\uparrow} \rangle = {\epsilon _{ijk} \over \sqrt{ 18}}
\left[ u_{i \downarrow} ^{\dag} d_{j \uparrow} ^{\dag} -
u_{i \uparrow} ^{\dag} d_{j \downarrow} ^{\dag} \right]
u_{k \uparrow} ^{\dag} \vert 0 \rangle
\end{equation}
where $u ^{\dag}, d^{\dag}$ are up and down quark creation operators,
the arrows represent spin directions, and the latin indices are
color degrees of freedom. The part inside the square brackets
can be taken as a spin zero ``particle'' (thus having no magnetic moment)
of charge $+e / 3$ which interacts with the remaining spin 1/2 up
quark. To get a rough estimate of the electromagnetic field angular
momentum contained in a system with a spin zero ``particle'' of charge
$+e /3$ and a spin 1/2 particle of charge $+2e/3$ we take the magnetic
dipole of the spin 1/2 particle to be aligned along the positive
z-axis and then the magnitude  of the ${\bf L}_{em}$ from Eq. 
(\ref{amdipole}) is
\begin{equation}
\label{lm}
\vert {\bf L}_{em} \vert = \sqrt{ {\bf L}_{em} \cdot {\bf L}_{em}} =
{e M \sin \theta  \over 3 R}
\end{equation}
where $\theta$ is the angle from the z-axis to ${\bf R}$. 
$M = (e_u g_u/ 4 m_u)$ is the magnitude of the
magnetic moment of the up quark, with $g_u$ being the
electromagnetic gyromagnetic ratio. Taking $\theta = \pi / 2$,
$g_u \approx 2$, $m_u \approx 2$ MeV $\approx 4 m_e$, $R =
0.5$ Fermi $= 0.5 \times 10 ^{-13}$ cm,  and inserting numbers in 
Eq. (\ref{lm}) (using the units of Ref. \cite{jackson} and putting
a factor of $\hbar / c$ back in ${\bf M}$ and a factor of $c$ in
the denominator of Eq. (\ref{lm})) yields $\vert {\bf L}_{em} \vert
= 0.3 (\hbar /2)$. In this estimate the electromagnetic 
field angular momentum is on the order of the fundamental unit of
angular momentum and therefore could conceivably contribute to
the total internal angular momentum of the nucleon. In doing this
rough estimate we have taken some of the numbers so as to favour
a large electromagnetic angular momentum. Probably the most 
questionable insert is in the mass of the up quark, where we have
taken the lower limit given in Ref. \cite{pdhb}, and we have used
the current mass rather than the constituent mass. Using the
constituent mass would tend to lower the value of the field angular
momentum by one or two orders of magnitude thus making it less likely
that the electromagnetic field angular momentum plays any significant
role in the internal angular momentum of the nucleon. However we
have also taken $g_u \approx 2$ (which is the order of magnitude
value that one gets from indirect fits to observed baryon magnetic
moments \cite{don}). If $g_u$ were larger than $2$ this
would help increase the electromagnetic field angular momentum. Also
if the separation between the constituents, $R$, inside the nucleon 
were smaller than our estimate this would tend to increase
the electromagnetic field angular momentum. As an
aside this is probably why this field angular momentum does not
play a significant role in atomic size systems despite the fact that such
systems certainly have both charges and magnetic dipole moments :
in atomic scale systems the separation, $R$, between the constituents
is five orders of magnitude larger which would decrease the field
angular momentum by five orders of magnitude. Finally, any sub-quark or
sub-lepton models, where the constituents had charges and magnetic
moments, would almost certainly have to take this electromagnetic
field angular momentum into account, since the experimental 
upper limit on quark and lepton sizes is several orders of magnitude
smaller than the fermi size scale of the nucleons.

An objection to this estimate is that we have treated the
dipole-charge field angular momentum in a semi-classical way.
In some cases this can give accurate results, as in the case
of the Thomson dipole where requiring that the field angular
momentum equal some integer multiple of $\hbar / 2$ \cite{saha}
yields the Dirac quantization condition. A more
careful analysis of the Thomson dipole \cite{lipkin} shows that 
the field angular momentum is not a proper quantum mechanical
angular momentum when considered by itself, since it does not
satisfy the standard commutator $[L_i , L_j] = i \epsilon_{ijk} L_k$.
The same holds for the dipole-charge system : the field angular
momentum given in Eq. (\ref{fangmom1}) does not, by itself, satisfy the
standard angular momentum commutator. It is straightforward, however,
to show that the total angular momentum of the dipole-charge
system
\begin{equation}
\label{totmom}
{\bf J} = ({\bf R \times D}) + {\bf L} _{em}
\end{equation}
does satisfy $[J_i , J_j] = i \epsilon _{ijk} J_k$. In Eq.
(\ref{totmom}), ${\bf D} = -i \nabla - e {\bf A}$ is the covariant
derivative. To show that ${\bf J}$ satisfies $[J_i , J_j] 
= i \epsilon _{ijk} J_k$ it is necessary to use the fact that
the covariant derivative operator does not commute with itself,
but rather $[D_i , D_j ] = i e \epsilon _{ijk} B_k$ \cite{lipkin},
with $B_k$ being the magnetic field. 

From the above one might conclude that although it is technically
correct to include the electromagnetic field angular momentum 
in the total internal angular momentum of the nucleon so that 
Eq. (\ref{spin}) becomes
\begin{equation}
\label{spin1}
{1 \over 2} = {1 \over 2} \Delta \Sigma + L_q + L_g + \Delta G
+ L_p + \Delta P
\end{equation}
(where the last two terms are the photon orbital and spin 
contributions respectively) 
there is only a marginal chance that this is the
source of the extra angular momentum. Even with this pessimistic
view it is possible that the color version of the above mechanism
could be invoked to explain the internal angular momentum of
the nucleon. In addition to electric charge, quarks carry a
color charge, which when coupled with the spin 1/2 nature of the
quarks produces a color magnetic moment. The gauge field part of
the QCD angular momentum can be written just as in Eq. (\ref{fangmom1})
except with ${\bf E} \rightarrow {\bf E}^a$ and ${\bf B} \rightarrow
{\bf B}^a$, where ${\bf E}^a$ and ${\bf B}^a$ are the color-electric
and color-magnetic fields respectively. As in Ref. \cite{ji} one
can write the QCD field angular momentum in the following form
\begin{equation}
\label{qcdam}
{\bf L} _{QCD} = {1 \over 4 \pi} \int {\bf r} \times ({\bf E}^a
\times {\bf B}^a) d^3 x
\end{equation}
(The total QCD angular momentum also includes terms for the
quark spin and quark orbital angular momentum \cite{ji} which
we have not written out explicitly here).
Except for the color index $a$ this looks identical to the 
electromagnetic field angular momentum. At this point
it is impossible to continue rigorously as in the electromagnetic 
case, since the strongly coupled, non-linear nature of QCD prevents
one from determining the ${\bf E}^a$ and ${\bf B}^a$
produced by the quarks at all energy scales. At a large enough 
energy scale, so that asymptotic freedom would allow one to treat
the QCD interaction perturbatively, the non-confining part of 
the strong interaction potential should look similiar
to the electromagnetic interaction potential except for 
the replacement of $\alpha _{em} \rightarrow \alpha _s$
({\it i.e.} the one-gluon exchange potential would be the QCD 
version of the Breit-Fermi potential of QED as in Refs. \cite{DeR} 
and \cite{don}). Thus under the assumption that one can deal with the
one-gluon and confining parts of the color interaction separately,
the perturbative part of the color electric and magnetic fields 
should have the following functional form
\begin{equation}
\label{ebqcd}
{\bf E}^a = {g T ^a {\bf r}' \over r^3} \; \; \; \; \; \; \; \; \; \;
{\bf B}^a = T^a \left({3 {\bf r} ({\bf r \cdot M}) \over r^5} -
{{\bf M} \over r^3} \right)
\end{equation}
where $T ^a$ are the standard SU(3) matrices, and
${\bf M} = g {\cal G}_c {\bf S} / 2 m$ is the color magnetic moment,
with ${\cal G}_c$ being the color equivalent of the electromagnetic
g-factor. Inserting these color electric and magnetic fields from Eq.
(\ref{ebqcd}) into Eq. (\ref{qcdam}) we find that the QCD version of
the electromagnetic angular momentum of Eq. (\ref{amdipole}) is
\begin{equation}
\label{qcdamdi}
{\bf L} _{QCD} = - {4 \over 3} \left( {g {\bf M} \over R} - {g {\bf R}
({\bf M} \cdot {\bf R} ) \over R^3} \right)
\end{equation}
This is similiar to Eq. (\ref{amdipole}) except for the replacement
$e \rightarrow g$ and the factor of $-4/3$. Interpreting the proton
state vector given in Eq. (\ref{proton}) as a color-anticolor bound 
state of a spin 0 and spin 1/2 particles one finds that the 
sum over the color index $a$ gives $\Sigma _{a=1} ^8 T^a _1 T^a _2 =
- 4/3$ (where $1$ and $2$ refer to the spin 0 and spin 1/2 particles).
The magnitude of this angular momentum is then
\begin{equation}
\label{qcdmag}
\vert {\bf L} _{QCD} \vert = { 4 g M sin \theta \over 3 R}
\end{equation}
where $M = g {\cal G}_c / 4 m_u$ is the magnitude of the color
magnetic moment of the up quark. We take $\theta = \pi / 2$,
and ${\cal G} _c \approx 2$ as in the previous electromagnetic
case. Taking $\alpha _s \approx 1 = 137 \alpha _{em}$ we find
that even if we use the constituent mass for $m_u \approx 137 \times
2$ MeV $= 274$ MeV ({\it i.e.} so the factors of 137 cancel between
the coupling and mass terms) that the magnitude of the QCD
field angular momentum comes to $\vert {\bf L}_{QCD} \vert = 1.2
(\hbar / 2)$ relative to the estimate for the electromagnetic
case. Within this tree-level approximation for the nonconfining
part of the QCD interaction we find that the color field 
angular momentum will contribute a significant amount to the 
total spin of the nucleon.

The main objection to the preceding calculation is the lack of
a firm theoretical justification for using basically an electromagnetic
form for the QCD interaction in a regime where such an approximation
is questionable. The use of electromagnetic analogies for 
low energy QCD is not unprecedented.
In Ref. \cite{DeR} and \cite{don} the mass splittings of QCD bound
states (such as the $\Delta ^+$ and proton, or the $\Sigma ^0$
and $\Lambda$) are explained in terms
of a color ``hyperfine'' spin-spin interaction. The results are
qualitatively and, to an extent, quantitatively  accurate. In any
case there is certainly some energy scale where the color version
of the electromagnetic dipole-charge field angular momentum should
be rigourously applicable, and one would expect it to contribute
more angular momentum than the electromagnetic effect, due to
the presence of group theoretical color factors and also since
$\alpha _s > \alpha _{em}$.

As in the electromagnetic case it can be shown that ${\bf
L}_{QCD}$ is not a good quantum mechanical angular momentum when
considered by itself since it does not satisfy the standard
commutation relationship $[L_{i (QCD)} , L _{j (QCD)}] = i
\epsilon _{ijk} L_{k(QCD)}$. It is only the sum of this color
field angular momentum with the contribution coming from the
quarks that satisfies the proper commutation relationships. 
This indicates that a more rigorous treatment of either the
electrodynamic or the chromodynamic systems should take this into
account rather than treating the terms separately. This 
also raises questions about discussing the terms in Eqs. (\ref{spin}) 
or (\ref{spin1}) separately. The 
difficulties in determining individual terms ({\it e.g.} $L_q$) in 
Eqs. (\ref{spin}) or (\ref{spin1}), may be connected with the fact
that some terms may not be proper quantum mechanical angular momenta 
when considered separately.

Finally, one possible way to distinguish the two cases ({\it i.e.}
photons or gluons carrying the extra angular momentum of the 
nucleon) would be to compare the spin structure of the proton 
with that of the neutron. The state vector for the neutron is
\begin{equation}
\label{neutron}
\vert n _{\uparrow} \rangle = {\epsilon_{ijk} \over \sqrt{18} }
\left[ d ^{\dag} _{i \uparrow} u ^{\dag} _{j \downarrow} -
d^{\dag} _{i \downarrow} u^{\dag} _{j \uparrow} \right]
d^{\dag} _{k \uparrow} \vert 0 \rangle
\end{equation}
As with the proton state vector of Eq. (\ref{proton}) one can view 
this as a bound state of a spin 0 and a spin 1/2 particle. Now,
however, the electric charges on the spin 0 and spin 1/2 particles
are $+1/3$ and $-1/3$ respectively, thereby reducing the electromagentic
angular momentum of the neutron by a factor of 2 relative to the
proton. If the color field angular momentum is the source of the
nucleon's extra angular momentum then both the neutron and the proton
should carry the same color field angular momentum, since from
the state vectors of Eqs. (\ref{proton}) (\ref{neutron}) both are
similiar color-anticolor bounds states. 

\section{Discussion and Conclusions}

In this paper we have proposed that the rarely discussed
electromagnetic field angular momentum of a magnetic dipole
and an electric charge may provide a portion of the nucleon's
internal angular momentum not accounted for by the
valence quarks. (One source that does briefly mention the 
dipole-charge system's field angular momentum is Ref. \cite{feynman}).
From a rough, semi-classical estimate we find that this electromagnetic
field angular momentum could contribute to
the nucleon's spin. This points out that at least from
a techincal standpoint Eq. (\ref{spin}) should be expanded to include
terms for the photons as in Eq. (\ref{spin1}).

Even if the electromagnetic field angular momentum turns out not
to be the source of any of nucleon's spin, the color version of
this field angular momentum can be postulated as the source of
the extra internal angular momentum. While this is equivalent to
the statement that some of the nucleon's spin comes from the
gluons ({\it i.e.} the $L_g + \Delta G$ terms in Eq. (\ref{spin})),
it gives a more concrete and detailed picture of how this may come 
about from the interaction of color charges with color magnetic 
dipoles. This picture also allows one to make definitive predictions
about certain states. For example, in this picture the $\pi ^+ =
(1 / \sqrt{6}) [u_{i \uparrow} ^{\dag} {\bar d} _{i \downarrow} ^{\dag}
- u_{i \downarrow} ^{\dag} {\bar d} _{i \uparrow} ^{\dag} ] \vert
0 \rangle$ should carry no net field angular momentum in either
the electromagnetic or gluonic fields, since the field contributions
from the two terms cancel one another. Also, for mesons and bayrons
with heavy quarks the field angular momentum should play less of a role,
due to the inverse dependence of the magnetic moment on the particle's 
mass. At large enough energy scales this electromagnetic picture
of a field angular momentum arising from the interaction of a
dipole and charge should also work for QCD due to asymptotic freedom.
At lower energy scales it is not possible to rigorously determine a 
functional form for the color electric (${\bf E} ^a$) and color magnetic
(${\bf B} ^a$) fields produced by the quarks, due to the nonperturbative
nature of the QCD interaction. However, there are cases, such
as the mass splittings of hadrons with similiar quark content,
where the electrodynamic analogy appears to work qualitatively and
quantitatively \cite{DeR} \cite{don} despite the lack of a firm theoretical
justification. If the same postulate is applied to the color version 
of the dipole-charge field angular momentum one finds that this
gluonic field angular momentum could contribution significantly to
the total spin of the nucleon.

\section{Acknowledgements} I would like to thank Hannalore Roscher
and Justin O'Neill for their help and encouragement during the
writing of this paper.

\end{document}